# Extitation of wakefield by a laser pulse in a metallic-density electron plasma

*Denys Bondar (1, 2), Vasyl Maslov (1, 2), Iryna Levchuk (1), Ivan Onishcenko (1)*
*((1) National Science Center "Kharkov Institute of Physics and Technology", Kharkov, Ukraine*
*(2) V.N. Karazin Kharkov National University, Kharkov, Ukraine)*

*vmaslov@kipt.kharkov.ua  bondar@kipt.kharkov.ua*

Since it is possible to form laser pulses with a frequency much larger than the frequency of visible light, Prof. T.Tajima proposed using such pulse to accelerate the particles at its injection into the crystal. Here, the wakefield excitation in the metallic-density plasma and the electron acceleration by laser pulse are simulated. The accelerating gradient has been obtained approximately 3TV/m. It is shown that, as in ordinary plasma, with time beam-plasma wakefield acceleration is added to laser wakefield acceleration.

## 1. INTRODUCTION

The accelerating gradients in conventional linear accelerators are currently limited for technical reasons to ~100 MV/m [1], partly due to breakdown. Plasma-based accelerators have the ability to sustain accelerating gradients which are several orders of magnitude greater than that obtained in conventional accelerators [1, 2]. As plasma in experiment is inhomogeneous and nonstationary and properties of wakefield changes at increase of its amplitude it is difficult to excite wakefield resonantly by a long sequence of electron bunches (see [3, 4]), to focus sequence (see [5-9]), to prepare sequence from long beam (see [10-12]) and to provide large transformer ratio (see [13-19]). In [4] the mechanism has been found and in [20-23] investigated of resonant plasma wakefield excitation by a nonresonant sequence of short electron bunches. Due to the rapid development of laser technology and physics [1, 2, 24–36] laser-plasma-based accelerators are of great interest now. Over the past decade, successful experiments on laser wakefield acceleration of charged particles in the plasma have confirmed the relevance of this acceleration [23–28, 32, 37]. Evidently, the large accelerating gradients in the laser plasma accelerators allow to reduce the size and to cut the cost of accelerators. Another important advantage of the laser-plasma accelerators is that they can produce short electron bunches with high energy [24]. The formation of electron bunches with small energy spread was demonstrated at intense laser–plasma interactions [38]. Electron self-injection in the wake-bubble, generated by an intense laser pulse in underdense plasma, has been studied by numerical simulations (see [29]). Processes of a self-injection of electrons and their acceleration have been experimentally studied in a laser-plasma accelerator [39].

The problem at laser wakefield acceleration is that laser pulse quickly destroyed because of its expansion. One way to solve this problem is the use of a capillary as a waveguide for laser pulse. The second way to solve this problem is to transfer its energy to the electron bunches which as drivers accelerate witness. A transition from a laser wakefield accelerator to plasma wakefield accelerator can occur in some cases at laser-plasma interaction [40].

With newly available compact laser technology [41] one can produce 100 PW-class laser pulses with a single-cycle duration on the femtosecond timescale. With a fs intense laser one can produce a coherent X-ray pulse. T.Tajima suggested [42] utilizing these coherent X-rays to drive the acceleration of particles. Such X-rays are focusable far beyond the diffraction limit of the original laser wavelength and when injected into a crystal interacts with a metallic-density electron plasma ideally suited for laser wakefield acceleration [42].

In [43-48] it has shown that at certain conditions the laser wakefield acceleration is added by a beam-plasma wakefield acceleration. In this paper laser wakefield acceleration in a plasma of metallic-density, the maximum accelerating gradient in such a new medium, the transition to the regime of joint laser wakefield acceleration and beam-plasma wakefield acceleration are numerically simulated.

## 2. PARAMETERS OF SIMULATION

Fully relativistic particle–in–cell simulation was carried out by UMKA 2D3V code [49]. A computational domain (x, y) has a rectangular shape. $\lambda$ is the laser pulse wavelength. The computational time interval is $\tau = 0.05$, the number of particles per cell is 8 and the total number of particles is $15.96 \times 10^6$. The simulation of considered case was carried out up to 800 laser periods. The period of the laser pulse is $t_0 = 2\pi/\omega_0$. The s-polarized laser pulse enters into uniform plasma. In $y$ direction, the boundary conditions for particles, electric and magnetic fields are periodic. The metallic plasma density is $n_0 = 1.8 \times 10^{22}$ cm$^{-3}$. The critical plasma density $n_c = m_e\omega_0^2/(4\pi e^2)$. The laser pulse is defined with a "cos$^2$" distribution in longitudinal direction. The pulse has a Gaussian profile in the transverse direction. The longitudinal and transverse dimensions of the laser pulse are selected to be shorter than the plasma wavelength. Full length at half maximum equals $2\lambda$ and full width at half maximum equals $8\lambda$. $a_0 = eE_{x0}/(m_e c\omega_0) = 4$, $E_{x0}$ is the electric field amplitude. Coordinates $x$ and $y$, time $t$, electric field amplitude $E_x$ and electron plasma density $n_0$ are given in dimensionless form in units of $\lambda$, $2\pi/\omega_0$, $m_e c\omega_0/(2\pi e)$, $m_e\omega_0^2/(16\pi^3 e^2)$.

## 3. RESULTS AND DISCUSSION

We first consider the wakefield excitation by two laser pulses. The wakefield bubbles have been formed. It has been shown that after time, smaller than approximately 175 laser periods the laser wakefield acceleration is added by plasma wakefield acceleration. In Fig. 1 one can see that 1st electron bunch, self-injected and accelerated after the 2nd laser pulse in the 3rd bubble, at this time became the driver. However, in the 3rd wake bubble new bunch is already self-injected by the maximum accelerating $E_x$.

At this time 1-st bunch in 1-st wake bubble is al-ready under the effect of an almost zero $E_x$, and after some time it will become a driver. I.e. in Fig. 1 one can see that 1-st bunch, self-injected and accelerated after 1st laser pulse in 1-st wake bubble, at this time is still accelerated, but already in a small $E_x$, because it almost reached the middle of bubble.

In Fig. 1 one can see that the accelerating field reaches the maximum value at the time $t=132t_0$ and this value is equal to $E_x^{arb.un.} = 0.13508$. In the dimensions this accelerating field is equal to $E_x^{V/m} = 2.75\,\text{TV/m}$.

Later, this new self-injected bunch is initially accelerated by three drivers: two laser pulses and the second electron bunch, and later by four drivers: two laser pulses, 1-st electron bunch in 1-st bubble and 1-st electron bunch in 3-rd bubble.

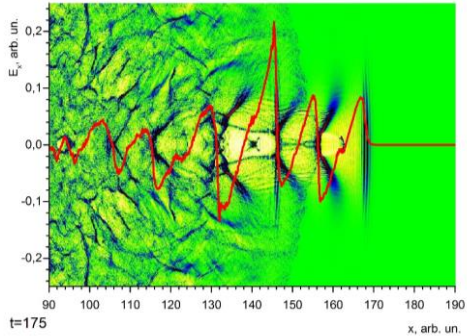

*Fig. 1. Wake perturbation of plasma electron density $n_e$ and longitudinal wakefield $E_x$ (red line) excited by two identical laser pulses at the time $t = 175t_0$*

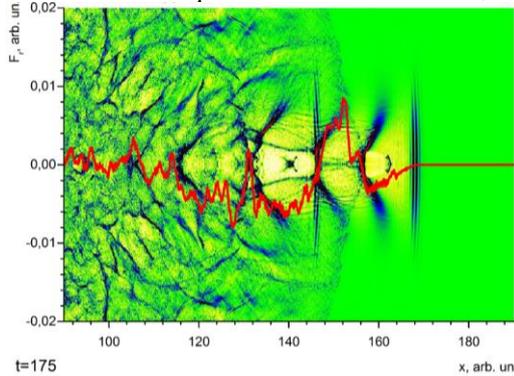

*Fig. 2. $n_e$ and off-axis transverse wake force $F_\perp \triangleright E_y - B_z$ (red) excited by two laser pulses at the time $t = 175t_0$*

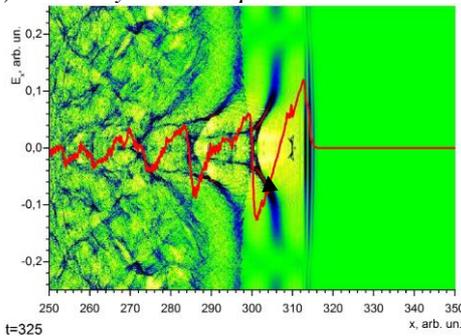

*Fig. 3. $n_e$, $E_x$ (red line) excited by laser pulse at $t=325t_0$*

In Fig. 2 one can see that the 1-st self-injected electron bunch in the 1-st wake bubble after the 1st laser pulse and the 1-st self-injected electron bunch in the 3-rd wake bubble after the 2nd laser pulse are under the effect of focusing wake force.

It is necessary to take into account the fact that after the 1st laser pulse in the 2nd bubble along its axis, a dense electron needle was formed, which provides a defocusing force along the 2nd bubble.

We now consider the wakefield excitation by one laser pulse. In Fig. 3 one can see that the 1st bunch self-injected and accelerated after the laser pulse in the 1-st wake bubble at this time already became a driver and it is decelerated by $E_x$. However, at the end of the 1st wake bubble a new bunch has already self-injected and it has moved towards the bubble in its middle.

In the case of wakefield excitation by one laser pulse, the maximum value of the accelerating wakefield in the normalized units is $E_x^{arb.un.} = 0.12641$. In the dimensions this accelerating wakefield is equal to $E_x^{V/m} \approx 2.57\,\text{TV/m}$.

From Fig. 4 one can see that 1-st and 2-nd self-injected electron bunches in 1-st wake bubble are under the effect of focusing wakefield.

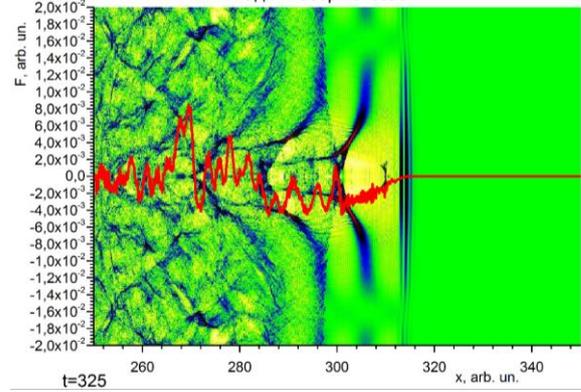

*Fig. 4. $n_e$ and $F_\perp \triangleright E_y - B_z$ (red line) excited by laser pulse at the time $t=325t_0$*

## 4. CONCLUSION

So, numerical simulation demonstrates the transition from the laser-wakefield acceleration in metallic-density plasma to extra beam-driven wakefield acceleration, providing additional acceleration of electron bunch. At interaction of two identical laser pulses with metallic-density plasma at some time 2nd self-injected witness-bunch in 3rd bubble is initially accelerated by three drivers: two laser pulses and the 1st electron bunch in 3rd bubble, and later by four drivers: two laser pulses, 1-st electron bunches in 1-st and 3-rd bubbles. At single laser interaction with metallic-density plasma at some time the accelerated 1st bunch in 1st bubble becomes a driver and it together with the partially dissipated laser pulse provides further acceleration of witness.

At wakefield excitation by the laser pulse or by two identical laser pulses in a metallic-density plasma, electrons are accelerated in electric fields which approximately equal several teravolts per meter.

It is necessary to investigate the effect of wakefield, excited due to internal bounded electrons (similar to wakefield in dielectric), on the investigated wakefield, excited due to free electrons of conductivity.